\documentclass[runningheads]{svmult}

\usepackage{makeidx}   
\usepackage{graphicx}  
\usepackage{subeqnar}  
\usepackage{multicol}  
\usepackage{physprbb}  
\makeindex             

\begin{document}
\def\apj{ApJ}
\def\araa{ARAA}
\def\aj{AJ}
\def\mnras{MNRAS}
\def\aap{A\&A}
\def\overarrow#1{\setbox0\hbox to 0pt{\hss$\scriptstyle\rightarrow$}%
                  #1\raise 1.6ex\box0}
\def\doverarrow#1{\setbox0\hbox to 0pt{\hss$\scriptstyle\rightarrow$}%
                  #1\kern.4ex\raise1.5ex\copy0\raise2.2ex\box0\kern-.4ex}
\def\hoverarrow#1{\setbox0\hbox to 0pt{\hss$\scriptstyle\rightarrow$}%
                 #1\kern.6ex\raise 1.6ex\box0\kern-.2ex}
\title*
{
Tales within Tales and Cutoffs within Cutoffs: 
\protect\newline What Sets the Mass Scale for Galaxies?}
\toctitle
{Tales within Tales: 
\protect\newline What Sets the Mass Scale for Galaxies?}
%
%
\titlerunning{The Mass Scale for Galaxies}
%
\author{Paul L. Schechter}
\authorrunning{Paul L* Schechter}
%
%
\institute{Massachusetts Institute of Technology, Cambridge MA 02139, USA}
\maketitle              

\begin{abstract} Please answer ``yes'' or ``no'':  \
\begin{enumerate}

\item Does the mass function for clusters of galaxies cut off 
exponentially?

\item Does the luminosity function for galaxies cut off 
exponentially?

\item Is the dependence of virial velocity on galaxy luminosity
a power law?  

\item Does the velocity function for galaxies cut off exponentially?
\end{enumerate}
\end{abstract}

\setcounter{section}{-1}
\section{Introduction}

The luminosities of cosmic lighthouses are limited by a variety of
physical processes.  The Eddington limit immediately comes to mind,
but like other such limits it is a function of mass.  The question of
limiting luminosities quickly becomes one of limiting masses.

At least four classes of cosmic lighthouses are on the program for
this meeting and each has a different typical mass associated with it.
It is my non-expert impression that for both stars and AGN we have some
understanding of the physics that limits their masses.  I will argue
that the mass scale for clusters of galaxies is readily explained, but
that our understanding (or at least my understanding) of the mass
scale for galaxies is very incomplete.

\section{The Mass Function for Clusters of Galaxies}

The mass scale for clusters of galaxies is set by cosmological
parameters which we take to be no different in character than the
fundamental microscopic constants.  The reigning paradigm is that
clusters of galaxies grow by gravitational instability from small
density perturbations in an otherwise uniform universe.

We start by asking how large a perturbation is needed at the time of
recombination to collapse into an object which has just virialized,
and find that we need a density perturbation with a fractional
amplitude of roughly 0.001.  We then ask how fluctuations in the
underlying dark matter distribution manifest themselves as observed
fluctuations in the cosmic microwave background.  We find the
corresponding fluctuations in the CMB on an angular scale which gives
a mass typical of of clusters of galaxies.

The typical mass for clusters of galaxies might manifest itself in
many different ways: it might be the mean of a Gaussian, the break in
two power laws, a cutoff in a single power law or something else.  The
typical mass sets the scale but it remains to determine the shape of
the mass function.  For the purpose of the present meeting, we
are especially interested the details of its high mass tail.

I did some work on the mass function for clusters as part of my PhD
thesis.  Gunn and Gott \cite{Gunn72} had just done their well known
work on the spherical collapse model for clusters.  It seemed to me
that since we knew that there were galaxies, whatever perturbations
gave rise to galaxies would, on a larger scale, give rise to clusters.
Starting, for the sake of argument, with the assumption that galaxies
were distributed in Poisson fashion I arrived at an expression for the
mass function for clusters of galaxies that consisted of a power law with
an exponential cutoff.

There were problems with the argument and they had me worried.  I had
equated two quantities that I knew where not the same thing.  The
nature of the swindle is more obvious if one describes the quantities in
words:
\begin{equation}
\left(
\begin{array}{c}
{\rm fraction~of}
\\
{\rm masses~} M
\\
{\rm with~} \delta > \delta_c
\end{array}
\right)
\approx
\left(
\begin{array}{c}
{\rm fraction~of~Universe}
\\
{\rm collapsed~in~objects}
\\
{\rm with~masses~} > M
\end{array}
\right) \quad .
\end{equation}
This {\it ansatz} made it possible to derive a mass function, but I
was very nervous about the slope at the low mass end.  An argument
that gave the same slope in any number of dimensions couldn't possibly
be right.  I struggled with this for the better part of a year.  Some
time thereafter Bill Press joined the effort.  He generalized my
Poisson result to a power-law spectrum of arbitrary index $n$,
emphasizing the self-similarity of the process and carried out what,
by Ed Bertschinger's \cite{Bertschinger98} reckoning, appears to have
been the first cosmological N-body simulations (with N = 1000) for the
purpose of checking our results.  We published what we had
\cite{Press74} and never did find a way to avoid the swindle.

Our mass function, 
\begin{equation}
n(M) =
{1 \over 2 \sqrt \pi}\left(1 + {n \over 3}\right)
{\overline\rho \over M^2} \left(M \over M*\right)^{3+n}
\exp\left[-\left(M \over M*\right)^{(3+n)/3}\right] \quad ,
\end{equation}
has undergone a number of extensions and improvements.  A fudge factor
has been successfully explained, and the approach has been ``extended"
(see Lacey and Cole \cite{Lacey93} and the review by
Schuecker {\it et al.} \cite{Schuecker01}) to include conditional
merger probabilities looking backward and forward in time.  It has
also been adapted to non-spherical collapse \cite{Sheth01}.  Its
deficiencies have been investigated by N-body experiment, leading
several groups to suggest modifications \cite{Sheth99,Jenkins01}.
Pierluigi Monaco writes \cite{Monaco98b}:
\smallskip

{\narrower
{\narrower
\noindent
The history of the mass function theory is reviewed in Monaco (1998)
\cite{Monaco98a} but it can effectively be summarized in a sentence:
there is a simple, effective and wrong way to describe the cosmological
mass function.  Wrong of course, does not refer to the results but 
to the whole procedure.
\par}
\par}
\smallskip
 
Not only did we get the calculation wrong, we had the input physics
wrong as well.  We assumed that the matter that was clustering was
strictly baryonic.  At recombination baryonic density perturbations
have roughly the same amplitude as the CMB fluctuations, which we now
know were much too small to produce today's clusters of galaxies.
What we didn't appreciate was that non-baryonic dark matter
perturbations, uncoupled to the photons and the baryons, could have
been growing while the baryonic perturbations were locked into the
photon fluid.  Clusters of galaxies are fundamentally
self-gravitating, pressure supported spheroids of dark, non-baryonic
matter.  Their baryons comprise only a small fraction of their mass.
And the galaxies from which they take their name include only a small
fraction of the baryons.

Such shortcomings notwithstanding, many investigators have found the
Press-Schechter recipe to be an acceptable first cut at the mass
function for clusters of galaxies.  The answer to our first question
appears to be ``yes'' -- the mass function has an exponential cutoff.

\section{The Luminosity Function for Galaxies}

\subsection{the shape of the luminosity function}

Hubble, Zwicky and George Abell all carried out studies of the
luminosity function for galaxies, and and all of them found that there
is a characteristic luminosity.  They differed considerably, however,
on the shape of the luminosity function.  Hubble found it to be
roughly Gaussian and Abell found the cumulative luminosity function to
be a broken power law.  

In one of the opening volleys of the science wars, Abraham Maslow
wrote \cite{Maslow66} wrote:
\smallskip

{\narrower
{\narrower
\noindent
I suppose it is tempting, if the only
tool you have is a hammer, to treat everything as if it were a nail.
\par}
\par}
\smallskip

\noindent 
It's true. My hammer was the power law with an exponential cutoff and
I hammered on the luminosity function for galaxies \cite{Schechter76}.

Many investigators have wielded this hammer since, and some have found
it unsatisfactory.  But it would still seem to be of some use.  In
their analysis of the SDSS commissioning data, Blanton {\it et al.}
find, to their evident surprise, that a power law with an exponential
cutoff fits their data better than an alternative, non-parametric
model \cite{Blanton01}.  The answer to the second question would also
appear to be ``yes.''

\subsection{implications for the mass function}

From the outset it seemed somewhat inconsistent, perhaps even
hypocritical, to use roughly the same functional form for both the
mass function for galaxies and the luminosity function for galaxies.
One might argue that the same process which gives rise to the cluster
mass function gives rise to the galaxy mass function at some earlier
time, but this argument is flawed.  The idea of self-similar
condensation was that the objects seen at one epoch would merge to
form larger objects.  How could they merge and yet survive?  We would
be eating our cake and having it too.  If the same process were to
work twice, an early epoch of structure had to be frozen in such a way
that it could survive subsequent mergers.  This question has been
addressed repeatedly, both theoretically and experimentally with
N-body simulations.  Only recently have the latter begun to give reliable
answers.  We discuss them in greater detail below, but for the moment
the important conclusion is that a small but significant fraction of
the mass, perhaps 10\%, survives in what look like galaxies.

\section{The Faber-Jackson and Tully-Fisher Relations}

In papers published within a year of each other, Faber and Jackson
\cite{Faber76} (henceforth FJ) and Tully and Fisher \cite{Tully77}
(henceforth TF) found, respectively, that elliptical and spiral
galaxies exhibited a similar scaling between internal velocities
(velocity dispersion and circular velocity) and galaxy luminosity.
They found luminosity varied roughly as the fourth power of observed
velocity, with a scatter about the mean relation of 30-40\%.  The FJ
relation has since been refined to include surface brightness by
Dressler {\it et al.} \cite{Dressler87} and Djorgovski and Davis
\cite{Djorgovski87} giving what the latter call the ``fundamental
plane.''

Bernardi {\it et al.} \cite{Bernardi01} recently completed a massive
study of the fundamental plane finding, among other things, that it
is, very nearly, a plane.  When they compute the mean velocity dispersion
at fixed absolute magnitude and correct for evolution they see little
if any deviation from a power law -- the FJ relation holds.  The
answer to our third question is another ``yes''.

\section{The Velocity Function for Galaxies}

\subsection{why not the mass function?}

As our subject is the mass scale for galaxies, the distribution of
virial velocities -- the ``velocity function'' -- may seem a
diversion, deflecting us from our goal.  Alas it is nearly impossible
to measure masses for galaxies.  Mass measurements combine a velocity
and a scale length measurement.  While one can measure scale lengths
for the baryons one sees in galaxies -- stars or gas -- there is every
reason to believe that this is not representative of the dark matter.
Baryonic collapse is dissipational while dark matter is
non-dissipational.  Differential collapse by roughly a factor of ten
is not unreasonable.  We exhaust the supply of stars and gas with
which to measure galaxy measure rotation curves and velocity
dispersions before we reach the outer limits of the dark matter.

By contrast, the velocities we measure in the baryonic matter {\it
are} representative of the dark matter.  The virial theorem, in its
simplest incarnation, tells us that the time averaged squared (virial)
velocity is given by

\begin{equation}
<v^2> = \left< \hoverarrow r \cdot {\partial \Phi \over 
\partial \hoverarrow r} \right> \quad ,
\end{equation}
where $\Phi$ is the gravitational potential.  The dark matter and the
stars orbit in the same potential, and would have the same virial
velocity on the same orbit.  Since galaxies appear to be nearly
isothermal, the righthand side of the above equation is not a strong
function of position, and it is not unreasonable to take the virial
velocity for the baryonic matter as a proxy for that of the dark
matter.

A similar circumstance holds for studies of clusters of galaxies,
which sometimes use ``temperature functions'' rather than mass
functions.  Masses for clusters are similarly dependent upon details
of how the scale length is measured.  By contrast measurement of the
temperature of X-ray emitting gas is relatively straightforward.

\subsection{the observations}

At first it would seem a straightforward matter to compute a ``velocity
function'' from the luminosity function and either the FJ or TF
relation, especially if the luminosity-velocity relations were indeed
power laws.  For example, if if the luminosity function cut off as
${exp}(-L/L*)$ and luminosity varied as $v^4,$ one would have
the velocity function cutting off as ${exp}[-(v/v*)^4]$.  If the luminosity-velocity
relation saturated at large velocities, the cutoff would be sharper.

But there are complications.  Bernardi {\it et al.}
\cite{Bernardi01} point out that in carrying out the above calculation,
one must take into account the spread in virial velocity at fixed
luminosity.  Moreover, there is a potential problem \cite{Kochanek01}
in covariances of the luminosity function and the velocity-luminosity
relation with hidden variables.  Pahre {\it et al}. (as reported by
Kochanek \cite{Kochanek01}) have taken such covariances into account.
Their velocity function, as shown in Kochanek's figures 6,7 and 9,
cuts off exceedingly sharply.

The separations of multiply imaged quasars vary as the square of the
virial velocity.  Kochanek \cite{Kochanek01} has computed a velocity
function from these, and while the statistics are meager, it is
consistent with an exponential cutoff.  Taking all of this into
consideration, the answer to our fourth question, as determined from
observations, is ``most probably.''

\subsection{the simulations}

Determining the merger histories for present day galaxies from
observations is a daunting problem, even for our own Milky Way.
Luckily the hierarchical condensation of dark matter can be simulated
with N-body experiments, at least in principal.  But for many years
N-body practitioners were frustrated by their simulations -- they
formed beautiful clusters but there were no galaxies.  Previous
generations of structure merged to form the present generation.
Sometime around 1986 this phenomenon was given a name by Jerry
Ostriker -- overmerging.  In the last two or three years several
groups (notably those based in New Mexico, Seattle and Garching)
appear to have traced the overmerging problem to the numerical
necessity of softening the $1/r$ gravitational potential.  It now
seems that substructure can survive to the present epoch.

This is a major accomplishment, and cause for celebration, but we
shouldn't forget that while some substructure survives, most of it is
destroyed \cite{Springel01}.  The astronomical jargon used to describe
this is still in flux, but the substructures which survive (identified
with today's galaxies) are called sub-halos and the larger structures
in which they are embedded (identified with today's clusters) are
called halos.  To order of magnitude, it appears that 10\% of the dark
matter remains more closely associated with sub-halos rather than the
larger halos.
\footnote{ This raises an interesting question question of physics
which, while not strictly applicable to the universe in which we live,
may nonetheless help to understand it.  The paper by Bill Press and
myself made much of the self-similarity of the growth of structure.
This depended first, upon the power-law nature of the perturbation
spectrum and second, on the assumption of an Einstein-de Sitter
cosmology.  Today we believe that the perturbation spectrum at
recombination is the product of a number of competing processes that
would have destroyed the power-law nature of any input spectrum.  But
we can still ask what would happen in a hypothetical Einstein-de Sitter
universe in which the input spectrum {\it was} a power law.  Would
self-similarity still hold?  In particular, would the fraction of mass
in sub-halos always be the same percentage of the total mass at all
times?  Or would we expect to find sub-sub-halos inside sub-halos and
so forth?  I suspect that this is the case, although I doubt anyone
has the code or the perseverance to carry this out.}

The next step after identifying sub-halos is to construct a mass
function.  Some investigators choose to construct a velocity
function as well, either for the sake of comparison with observation or
because it is less sensitive to the details of the algorithm used to
identify sub-halos.

While the results are relatively new and the details may change as the
experiments improve, two important conclusions can already be drawn.
First, it appears that the mass function for sub-halos is very much
steeper than the luminosity function for galaxies.  Second, there is
at best weak evidence for an exponential cutoff in the masses of
sub-halos -- the mass functions look very much like power laws.

The same holds true for velocity functions.  They look like power laws
without any obvious cutoff \cite{Bullock01,Springel01}.  The
answer to our fourth question, as determined from N-body simulations,
is ``evidently not.''  In an odd way this is reassuring -- it is not
obvious what would introduce either a mass scale or velocity scale for
the sub-halos, so it is just as well that we don't see one.

We face a serious dilemma.  Observation gives a positive answer
to our fourth question and N-body experiment gives a negative
answer.

\section{The Baryonic Mass Function for Galaxies}

As discussed in the previous section, the dark matter mass function
for galaxies is perversely difficult to measure.  By comparison the
baryonic mass function is fairly easy -- indeed to first order it
is the luminosity function scaled by a mass-to-light ratio.

The N-body experiments in which sub-halos form within halos involve
only dark matter particles -- they interact only by non-dissipative,
mutual gravitation.  By contrast, luminosity functions, both optical
(for galaxies) and X-ray (for clusters), describe baryonic matter,
which can radiate away its energy.  All manner of physics can cause
dark matter and baryonic matter, however uniformly distributed at
first, to become separated over the course of time.  There is every
reason to think that mass function for baryons will therefore be quite
different from the mass function for sub-halos (and halos).

A variety of ``gastrophysical'' mechanisms have been invoked to
explain the differences between the observed (baryonic) luminosity
function for galaxies and sub-halo mass function derived from N-body
experiments.  At the low mass end, baryons must either be ejected from
sub-halos or something must prevent them from collapsing along with the
collapse of dark matter.  Somerville \cite{Somerville01} has used the
word ``squelching'' to describe the suppression of dwarf galaxy
formation by the re-ionization of the IGM.  There may be some
disagreement about the physics but it's an apt word to describe the
phenomenon.

Whatever causes the flattening of the mass function at the low mass
end, something else must cause a steepening at the high mass end.
White and Rees \cite{White78} proposed that the cooling time for
baryonic gas, assumed to be uniformly distributed in a dark matter
potential well, increases with increasing mass to the point at
which the baryons no longer have time to cool and condense.  It should
be remembered, however, that White and Rees went to considerable
lengths to emphasize that substructure would be destroyed.  If
substructure is not destroyed, then the gas associated with this
substructure would be clumpy and would have much shorter cooling
times.

Whatever the detailed physics, the underlying idea of White and Rees
would seem to be that sub-halos with large internal velocities exist,
but we don't see them because their baryons could not and have not
condensed.

A half-dozen groups are engaged in a heroic enterprise that goes by
the name ``semi-analytic modeling."  They attempt to form galaxies in
dark matter halos, taking care to include the gravitational
condensation of halos and the gas-dynamical condensation of baryons
within these halos.  They include star formation, energy feedback,
chemical enrichment and obscuration by dust.  All of these are
accomplished by a series of more-or-less realistic recipes.  More
recently some of these modelers have substituted high resolution
N-body simulations for analytically derived halo histories.  These
models have many free parameters -- enough to fit almost any set of
constraints.  Yet with all these parameters, the modelers still have
difficulty in cutting off the bright end of the luminosity function.  The
adopted values of the parameters seem extreme, and in some cases the
assumptions seem {\it ad-hoc}.

The TF and FJ relations place strong constraints on the recipes.  One
sort of recipe might have the fraction of the baryons which condense
within a halo decrease as the halo velocity dispersion increases.
A second sort might have the condensation of baryons in a halo be an
all-or-nothing affair, but with baryon condensation increasingly
unlikely with increasing halo dispersion.  Recipes of the first sort
might run afoul of the velocity-luminosity constraints, producing
halos with large velocities but relatively low luminosities.

I suspect that something on the order of the solution proposed by
White and Rees will ultimately prove successful in inhibiting the
collapse of baryons into the largest halos, thereby producing a
baryonic mass cutoff for galaxies.  But for the moment the modelers
have not yet persuaded each other; one wonders whether they have even
persuaded themselves.

For the sake of argument let's grant that something squelches the
condensation of baryons into low mass halos that something else
inhibits the condensation of baryons into high mass halos.  This
brings us back to the velocity function constructed from gravitational
lensing.  If baryonic infall had no effect on the structure of
sub-halos, the distribution of lens separations could be derived from
the velocity function determined from N-body simulations and one ought
not to see a cutoff.  But White and Kochanek \cite{White01} have
argued that baryonic infall {\it does} alter the gravitational
potentials of halos, particularly at the relatively small radii
sampled by lensed quasars, so even the lensed quasar separation
distribution may exhibit a cutoff, or at least a kink.  While this
might solve our problem, it should be noted (as they do) that their
model reproduces the observed distribution of lens separations only if
the baryonic mass fraction in galaxies is larger than is usually
assumed.

\section{cD Galaxies: Sub-Halos or Halos?}

No discussion of galaxies as lighthouses would be complete without
mention of cD galaxies.  It is helpful to make a distinction between
those which are instantaneously the brightest and those which are the
brightest when averaged over cosmic timescales.  In their
contributions to these proceedings {\sc Longair} and {\sc Cesarsky}
discuss sub-mm sources and ULIRGs, which are instantaneously the
brightest galaxies in the universe.  Averaged over time this honor may
belong, instead, to the cD galaxies found at the centers of many (but
not all) clusters of galaxies.

The ``cD'' classification originated in a paper by Matthews, Morgan
and Schmidt \cite{Matthews64} (henceforth MMS).  The lowercase ``c"
comes from Morgan's classification for supergiant stars and the
uppercase D stands for diffuse.  With NGC 6166 in Abell 2199 as the
archetype \cite{Morgan65}, these galaxies had large, low surface
brightness halos.  They frequently seemed to harbor radio sources.

The designation cD is sometimes taken to stand for the words central
and dominant.  While this was not the original meaning of MMS, there
is more than a bit of truth in it.  Such galaxies are indeed found
preferentially at the centers of clusters of galaxies and they tend to
be the dominant galaxy.  Oegerle and Hill \cite{Oegerle94} have found
that they also tend to lie at the center of the velocity distribution
for the cluster in which they reside.

There is an old but still interesting question as to whether the cD
galaxies are members of the same population as the other galaxies in a
cluster, or are in some way special.  If the luminosities of galaxies
really do cut off exponentially, there's no explaining cD's.

There are other peculiarities associated with cD's as well -- their
velocity dispersion profiles rise rather than fall as a function of
radius.  This is consistent with their extended envelopes and even
with a flattening of the slopes of their light profiles.  Moreover
they seem to have an excess of globular clusters.

The same N-body experiments which give rise to interesting
substructure also seem to produce a condensation at the center of each
cluster which has the density of a galaxy.  The special position of
galaxies at the centers of cluster potentials might make them
different.  Today they seem largely to be star piles, the piling up of
fragments of previously formed galaxies.

At earlier times cD's may have been special in a different way.
The galaxies at the centers of potentials are the natural sinks for
cooling baryons, and one expects baryons to collect there -- one would
think in disks.  In the semi-analytic models the galaxy at the center
of a halo occupies a unique position -- it is the only galaxy which
accretes baryons.  In such a model a dark matter halo collects baryons
only until it is subsumed into a larger halo in which it is no longer
in the privileged central position.  The other galaxies in the halo
are called satellites to distinguish them from the central galaxy.
But if they have any baryons, then at one time they must have been at
the center of the cluster.  In short, every galaxy must have at one
time been a central dominant galaxy, though not a cD galaxy in the
sense of MMS.

\section{Conclusion}
The answers to the four questions posed in the abstract are yes, yes,
yes and yes-and-no.  The observed velocity function for galaxies
appears to cut off exponentially while the velocity function for
N-body sub-halos appears not to.  There is as yet no detailed physical
model which cuts off the condensation of baryons into dark matter
halos in a manner which conforms to the observations.

%

\end{document}